\documentclass[11pt,letterpaper]{article}

\usepackage[T1]{fontenc}
\usepackage[utf8]{inputenc}
\usepackage[margin=1in]{geometry}
\usepackage{amsmath,amssymb}
\usepackage{graphicx}
\usepackage{booktabs}
\usepackage{authblk}
\usepackage{cite}
\usepackage{placeins}
\usepackage[hidelinks]{hyperref}

\title{
Calibration-Free Photonic Memory in a
Silicon Photonics Platform
}

\author[1]{Ankur Singh%
\thanks{Corresponding author:
\href{mailto:ankur.singh@wisc.edu}
{ankur.singh@wisc.edu}}}

\author[1]{Akhilesh Jaiswal}

\affil[1]{
University of Wisconsin--Madison,
Madison, WI 53706, USA
}

\date{}

\begin{document}

\maketitle

\begin{abstract}
We propose a calibration-free photonic memory bitcell
based on nonresonant variable optical attenuators in
the GlobalFoundries 45SPCLO platform. The cell maintains
complementary electrical states through electro-optical
feedback without cavity-resonance alignment.
Wordline-controlled attenuators enable selective writing
through differential optical bitlines, while optical
readout shares the holding input and output splitters,
avoiding additional read attenuators or a separate laser
source. Circuit simulations demonstrate writing with
$10~\mathrm{ps}$ optical pulses at $2~\mathrm{mW}$
peak power. The cell retains data as long as optical
and electrical power are supplied, with retention
simulated for $5~\mathrm{s}$ per logic state.
The evaluated transition consumes an incremental
write energy of $0.562~\mathrm{pJ}$, including circuit
electrical energy and the laser contribution at
20\% wall-plug efficiency. The bitcell layout occupies
$147.92\times219.93~\mu\mathrm{m}^{2}$.
\end{abstract}

\section{Introduction}

Integrated photonics has advanced rapidly in data
communication, computing, and sensing, while data
storage is still mainly performed using electronic
memory~\cite{glick2023integrated,kaiser2025opticam,
mao2019photonic,pintus2025integrated}. Storing incoming
optical data in electronic memory requires
photodetection and electrical data transfer, while
retrieval into an optical link requires electro-optic
modulation. Depending on the interface architecture,
analog-to-digital and digital-to-analog conversion
may also be required~\cite{lam2026neuromorphic}.
These operations introduce energy and latency
overheads. For the interface considered
in~\cite{ashtiani2025programmable}, the estimated
electronic data-transfer and conversion overhead
is at least $3.5~\mathrm{pJ/bit}$.
Integrated optical memories have therefore been
investigated extensively during the past two
decades~\cite{alexoudi2020optical}.

Static optical memories have been demonstrated
using semiconductor optical amplifiers in
Mach--Zehnder interferometer
configurations~\cite{pleros2008optical,pitris2016wdm}
and through cross-gain
modulation~\cite{fitsios2012dual,vagionas2012optical}.
These approaches can require substantial bias
currents and relatively large footprints.
Coupled microring and microdisk lasers provide
alternative bistable storage mechanisms, but require
active optical gain and material integration beyond
standard silicon photonic
components~\cite{hill2004fast,trita2013monolithic,
liu2010ultra}. Photonic-crystal nanocavities achieve
femtojoule-level switching
energies~\cite{nozaki2012ultralow}, although their
operation depends on cavity dynamics and the
associated material system. Phase-change memories
embedded in waveguides provide nonvolatile,
multilevel storage. Rios et
al.~\cite{rios2015integrated} demonstrate writing
with pulses as short as $10~\mathrm{ns}$, while
their multilevel demonstration uses
$100~\mathrm{ns}$ write pulses.

More recently, optical memory cells have been
implemented using components available in commercial
silicon-photonics foundries. A cross-coupled
differential photonic latch has been experimentally
demonstrated on the GlobalFoundries 45SPCLO platform,
with simulated writing using $50~\mathrm{ps}$
optical pulses at
$1~\mathrm{mW}$~\cite{abdullah2025demonstration}.
A programmable photonic processor has also been
used to characterize microring-based universal
optical logic gates and simulate a set-reset
latch~\cite{ashtiani2025programmable}.
Both architectures employ resonant devices.
The demonstrated photonic latch is evaluated
through simulations with input-wavelength offsets
of $\pm25~\mathrm{pm}$ and transient electrical
disturbances~\cite{abdullah2025demonstration}.
Related pSRAM simulations compensate for process
and temperature variations by retuning the optical
source wavelength~\cite{kaiser2026design}.
Maintaining alignment between the optical source
and the resonant devices consequently remains
an array-level design consideration.

In this work, we introduce photonic VOA-based
optical memory (pVOM) using components available
in the GlobalFoundries 45SPCLO platform.
The cell uses nonresonant variable optical
attenuators (VOAs) in an electro-optical feedback
loop to maintain complementary electrical states,
enabling optical writing and readout without
cavity-resonance alignment. Circuit simulations
demonstrate writing with $10~\mathrm{ps}$ optical
pulses at a peak power of $2~\mathrm{mW}$,
with an incremental write energy of
$0.562~\mathrm{pJ}$ for the evaluated transition.
Wordline-controlled access VOAs extend the cell
to an array, with selective writing evaluated
through simulation of a $4\times4$ memory array.

\section{Proposed VOA-Based Optical Memory}

\begin{figure}[htbp]
    \centering
    \includegraphics[width=\linewidth]
    {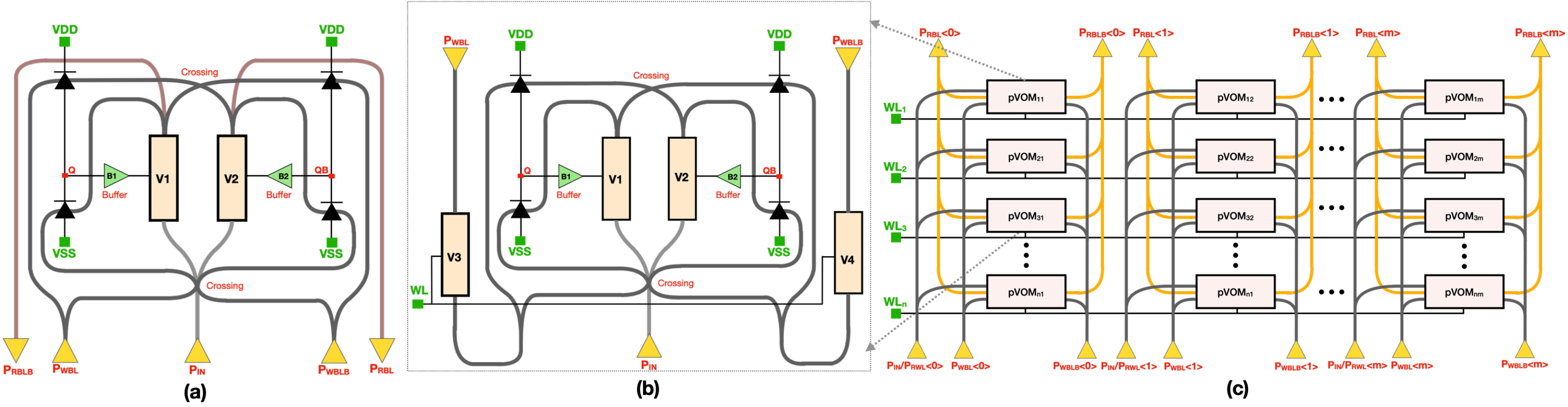}
    \caption{
    Proposed pVOM architecture:
    (a) differential VOA-based latch with
    complementary optical readout,
    (b) wordline-controlled memory cell with
    access VOAs $V_3$ and $V_4$, and
    (c) row--column memory array with shared
    wordlines and optical bitlines.
    }
    \label{fig:pvom_cell}
\end{figure}

Fig.~\ref{fig:pvom_cell}(a) shows the proposed
VOA-based photonic memory latch with optical
readout. The storage core consists of two VOAs
($V_1$ and $V_2$), photodiode branches, and buffers
$B_1$ and $B_2$, forming an electro-optical
feedback loop. The stored data is represented
by the complementary voltages at nodes $Q$ and
$\overline{Q}$. These voltages control the
attenuation states of the VOAs through the
buffers, while the transmitted optical power
determines the photodiode response in the
feedback branches.

The optical input $P_{\mathrm{IN}}$ supplies
the latch, and the differential inputs
$P_{\mathrm{WBL}}$ and $P_{\mathrm{WBLB}}$
provide optical write access. The stored state
is available through the complementary optical
read outputs $P_{\mathrm{RBL}}$ and
$P_{\mathrm{RBLB}}$.

\subsection{Hold Operation}

During hold operation, both optical write inputs
remain inactive, while $P_{\mathrm{IN}}$ and
the electrical supply remain applied.
For a stored state with $Q$ high and
$\overline{Q}$ low, the complementary node
voltages establish different attenuation levels
in $V_1$ and $V_2$. The resulting optical powers
produce a photodiode response that reinforces
the voltage difference between the storage nodes.

The buffers couple these node voltages to the
VOA control terminals, completing the regenerative
feedback loop. The opposite stored state is
maintained through the corresponding reversal
of the branch conditions. After the write pulse
is removed, the latch therefore retains the
selected state without repeated write pulses,
as long as optical and electrical power remain
supplied. Since the storage core uses nonresonant
VOAs, the feedback does not require alignment
of the input wavelength to a cavity resonance.

\subsection{Write Operation}

Fig.~\ref{fig:pvom_cell}(b) shows the memory cell
with wordline-controlled write access for array
operation. Two additional VOAs, $V_3$ and $V_4$,
are inserted in the $P_{\mathrm{WBL}}$ and
$P_{\mathrm{WBLB}}$ paths, respectively.
Both access VOAs are controlled by a common
electrical wordline (WL).

When $\mathrm{WL}=0$, the access VOAs operate
in their low-attenuation state, allowing the
differential optical write signals to reach
the latch. An optical pulse applied to one
write bitline changes the photocurrent balance
between the feedback branches and drives the
storage nodes toward the selected complementary
state. As the node voltages change, the
transmission levels of $V_1$ and $V_2$ change
accordingly, and the regenerative feedback
establishes the new state.

Applying a pulse to the opposite write bitline
reverses the stored data. The write pulse must
provide sufficient optical power and duration
to overcome the previously retained state.
Once the new state is established, the feedback
maintains the stored node levels after the
write pulse is removed.

When $\mathrm{WL}=1$, $V_3$ and $V_4$ enter
their high-attenuation state, suppressing the
optical write signals and returning the cell
to hold operation.
Fig.~\ref{fig:pvom_cell}(c) shows the proposed
memory array. Cells in each row share a common
electrical wordline, while each column shares
a differential optical write-bitline pair.
Selecting a row enables its access VOAs to
pass the optical write signals to the latches,
allowing the cells in that row to be written
in parallel. In unselected rows, the access
VOAs attenuate the write signals. Their
residual transmission must remain below the
write-disturb threshold to preserve the
stored states.

\subsection{Read Operation}

The optical input port $P_{\mathrm{IN}}$
is used for both maintaining the latch and
reading the stored data. As shown in
Fig.~\ref{fig:pvom_cell}(a), an optical splitter
at the output of each VOA, $V_1$ and $V_2$,
divides the transmitted power into two paths:
one sustains the latch feedback, while the
other provides the corresponding optical read
output, $P_{\mathrm{RBL}}$ or
$P_{\mathrm{RBLB}}$.

During read operation, the power applied at
$P_{\mathrm{IN}}$ is increased above the level
used for hold operation, while the write
inputs remain inactive. The increased input
power provides higher optical power at the
read outputs, whose levels are determined by
the VOA transmission states associated with
$Q$ and $\overline{Q}$.

The stored bit is therefore read through the
existing optical paths without additional
read VOAs or a separate laser source.
After readout, $P_{\mathrm{IN}}$ is returned
to the holding-power level. The hold and read
input powers, complementary output levels,
and preservation of the stored state during
readout are evaluated in the following section.

\section{Simulation Results and Discussion}

The proposed pVOM is evaluated through circuit
simulations in Cadence Virtuoso using the
GlobalFoundries 45SPCLO silicon-photonics platform.
The simulations use a supply voltage of
$1.8~\mathrm{V}$ and an operating wavelength
of $1310~\mathrm{nm}$, with a holding input
power of $50~\mu\mathrm{W}$ and a peak optical
write power of $2~\mathrm{mW}$.
The complementary storage-node voltages and
optical outputs are monitored to evaluate
write, retention, and read operation.

\begin{figure}[htbp]
    \centering
    \includegraphics[width=\linewidth]
    {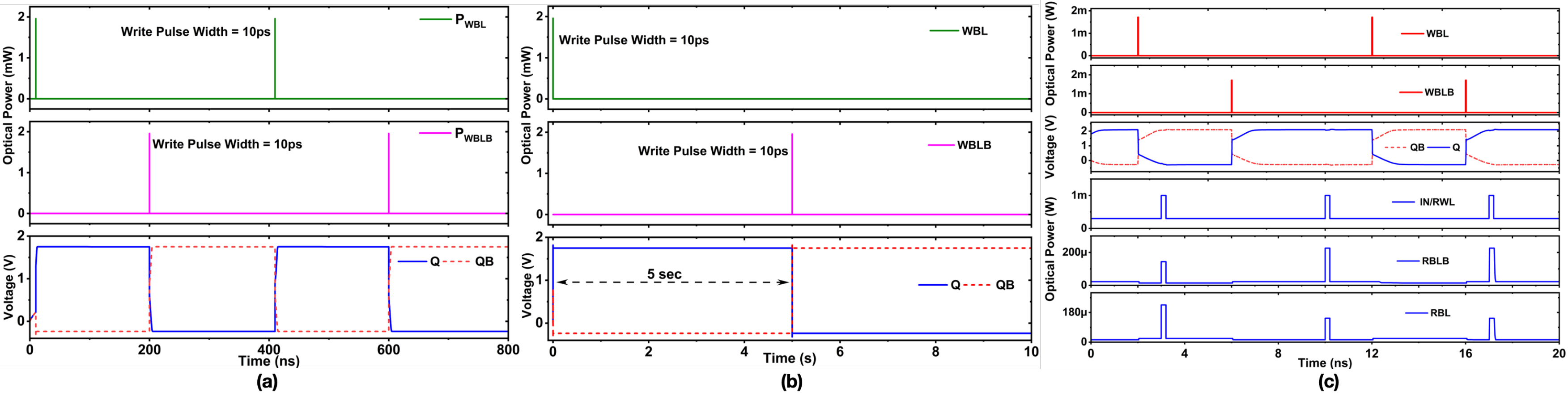}
    \caption{
    Simulated pVOM operation:
    (a) differential writing with
    $10~\mathrm{ps}$ optical pulses,
    (b) static retention simulated for
    $5~\mathrm{s}$ per logic state, and
    (c) optical readout through the shared
    holding input without changing the
    stored logic state.
    }
    \label{fig:write_retention}
\end{figure}

\subsection{Write Operation}

Fig.~\ref{fig:write_retention}(a) shows the
transient response under alternating optical
write pulses with a peak power of approximately
$2~\mathrm{mW}$. A pulse on
$P_{\mathrm{WBL}}$ establishes a high voltage
at $Q$ and a low voltage at $\overline{Q}$,
whereas a pulse on $P_{\mathrm{WBLB}}$
reverses the stored state.

The storage nodes settle at approximately
$1.8~\mathrm{V}$ and $-0.2~\mathrm{V}$
for the high and low states, respectively.
After each write pulse is removed, the
complementary voltages remain at their
retained levels until the opposite write
input is applied. The repeated state reversals
over the $800~\mathrm{ns}$ simulation window
demonstrate differential optical writing
followed by hold operation.

\subsection{Static Data Retention}

Fig.~\ref{fig:write_retention}(b) shows the
retention response over a $10~\mathrm{s}$
simulation window. An initial pulse on
$P_{\mathrm{WBL}}$ establishes $Q$ high
and $\overline{Q}$ low. Both write inputs
then remain inactive, and the latch maintains
this state until approximately
$t=5~\mathrm{s}$.

A pulse on $P_{\mathrm{WBLB}}$ subsequently
reverses the stored data, which is retained
for the remainder of the simulation.
Each state is therefore maintained for
approximately $5~\mathrm{s}$ without
additional write pulses. No significant
drift is visible at the plotted scale.
The cell operates as a static memory,
retaining its state while the electrical
supply and holding illumination remain applied.

\subsection{Read Operation}

Fig.~\ref{fig:write_retention}(c) shows the
read response when $P_{\mathrm{IN}}$ is
increased from the holding level of
$50~\mu\mathrm{W}$ to $1~\mathrm{mW}$,
with both write inputs inactive.
The splitters at the outputs of $V_1$
and $V_2$ direct part of the transmitted
power to the read ports, while the remaining
power sustains the latch feedback.

The complementary read outputs reach
$236$ and $143~\mu\mathrm{W}$ for the
stored state $Q=1$, and their levels
interchange when the opposite state is
stored. The resulting read extinction ratio
is $2.18~\mathrm{dB}$, calculated as
\begin{equation}
\mathrm{ER}
=
10\log_{10}
\left(
\frac{P_{\mathrm{H}}}{P_{\mathrm{L}}}
\right),
\end{equation}
where $P_{\mathrm{H}}$ and $P_{\mathrm{L}}$
denote the high and low read-output powers.
The storage-node voltages exhibit transient
variations during readout, but the complementary
logic states remain unchanged after the read pulse.

The write, retention, and read operations are
also tested at several O-band wavelengths
with the electrical bias and optical powers
held constant. At each tested wavelength,
the cell exhibits responses similar to those
shown in Fig.~\ref{fig:write_retention}(a)--(c),
without requiring resonance tuning.
Here, calibration-free operation refers
specifically to the absence of resonance
alignment in the storage mechanism.
The cell still requires suitable electrical
bias and optical power to sustain feedback.

\subsection{Energy and Physical Implementation}

The switching energy of the proposed memory
includes the electrical energy consumed by
the circuit and the energy required to
supply the optical write pulse.
The instantaneous electrical power is
obtained from
$P_{\mathrm{elec}}(t)=V(t)i(t)$.
To exclude the static holding contribution,
the quiescent hold power is subtracted
before integration, giving
\begin{equation}
E_{\mathrm{sw,elec}}
=
\int_{t_1}^{t_2}
\left[
P_{\mathrm{elec}}(t)
-
P_{\mathrm{hold,elec}}
\right]dt.
\label{eq:electrical_energy}
\end{equation}

The integration window begins
$10~\mathrm{ps}$ before the optical write
pulse and extends until $Q$ and
$\overline{Q}$ have reached their
steady-state values.
For a rectangular $2~\mathrm{mW}$,
$10~\mathrm{ps}$ optical write pulse,
the incident optical energy is
\begin{equation}
E_{\mathrm{opt}}
=
P_{\mathrm{W}}t_{\mathrm{W}}
=
20~\mathrm{fJ}.
\label{eq:optical_energy}
\end{equation}

Assuming a laser wall-plug efficiency
of $\eta=0.2$, the corresponding
source-referred laser energy is
\begin{equation}
E_{\mathrm{laser}}
=
\frac{E_{\mathrm{opt}}}{\eta}
=
100~\mathrm{fJ}.
\label{eq:laser_energy}
\end{equation}

The incremental write energy is therefore
\begin{equation}
E_{\mathrm{write}}
=
E_{\mathrm{sw,elec}}
+
E_{\mathrm{laser}}.
\label{eq:write_energy}
\end{equation}

For the evaluated transition, the electrical
switching energy is $461.6~\mathrm{fJ}$.
Including the $100~\mathrm{fJ}$
source-referred laser contribution gives
an incremental write energy of
$0.562~\mathrm{pJ}$.
This value excludes continuous holding
energy, which increases with retention
duration according to
\begin{equation}
E_{\mathrm{hold}}
=
P_{\mathrm{hold}}T_{\mathrm{hold}}.
\label{eq:hold_energy}
\end{equation}
Here, $P_{\mathrm{hold}}$ includes the
electrical holding power and the laser
power required to supply the holding
illumination, accounting for wall-plug
efficiency.

\begin{figure}[htbp]
    \centering
    \includegraphics[width=0.60\linewidth]
    {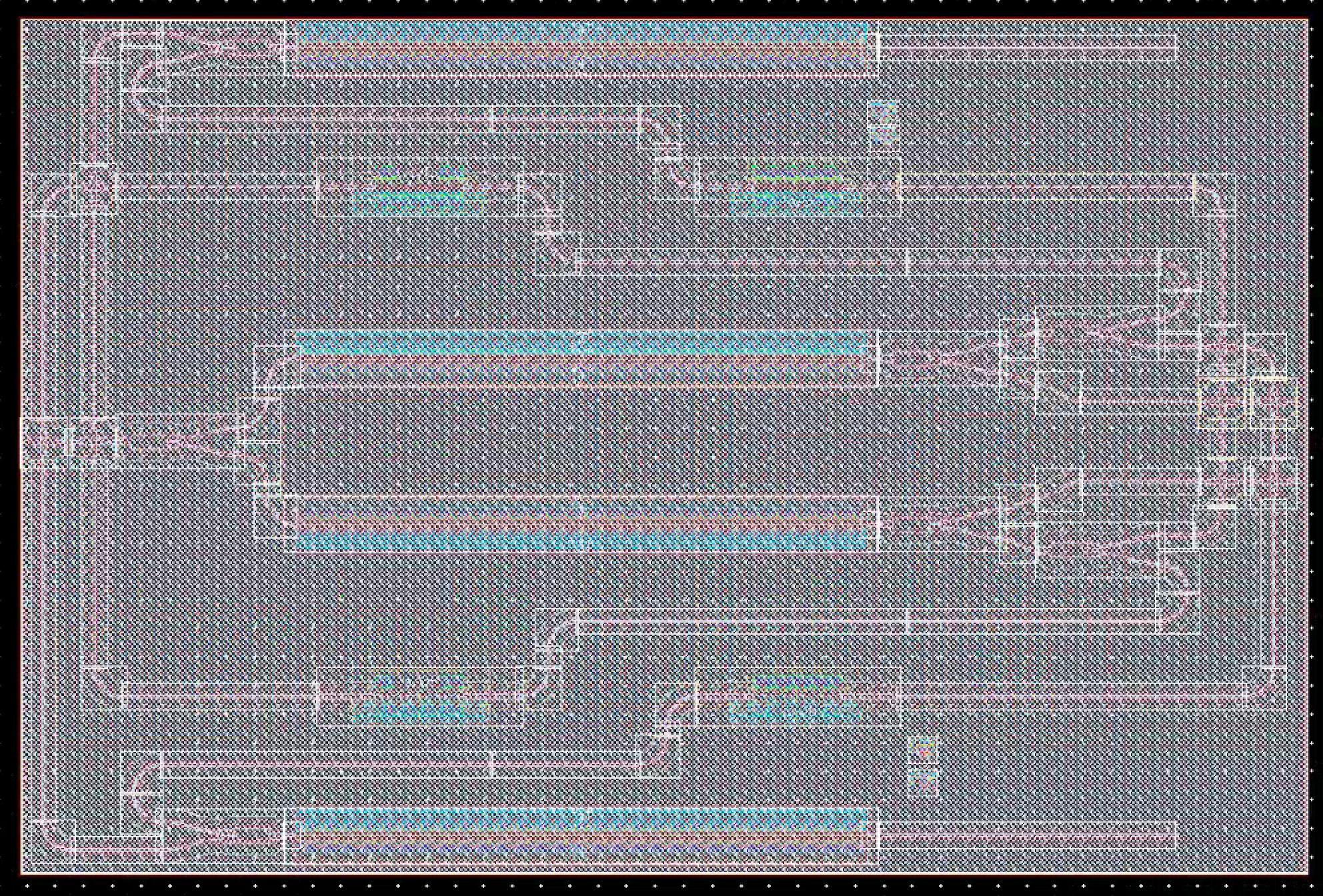}
    \caption{
    Physical layout of the proposed pVOM
    bitcell occupying
    $147.92\times219.93~\mu\mathrm{m}^{2}$.
    }
    \label{fig:layout}
\end{figure}

Fig.~\ref{fig:layout} shows the proposed
bitcell layout, occupying
$147.92\times219.93~\mu\mathrm{m}^{2}$
($0.0325~\mathrm{mm}^{2}$).
The area and energy of array interconnects
and peripheral circuits require separate
evaluation.

\subsection{Comparison and Discussion}

Table~\ref{tab:memory_comparison} compares
the proposed pVOM with previously reported
photonic memories.
The cross-coupled microring pLatch
in~\cite{abdullah2025demonstration}
experimentally demonstrates optical writing,
readout, and powered retention, with simulated
writing using $50~\mathrm{ps}$ optical pulses
at $1~\mathrm{mW}$.
Its reported switching energy and footprint
are $0.5~\mathrm{pJ}$ and
$0.1~\mathrm{mm}^{2}$, respectively.

The proposed pVOM supports simulated writing
with $10~\mathrm{ps}$ optical pulses
at $2~\mathrm{mW}$ and occupies
$0.0325~\mathrm{mm}^{2}$, approximately
32.5\% of that reported footprint.
The area comparison depends on the access,
readout, and routing components included
in each implementation.
Both architectures employ differential
electro-optical feedback, while pVOM uses
nonresonant VOA attenuation to establish
the stored states.

Resonance tuning can introduce additional
power overhead in microring-based systems.
For example,
Ref.~\cite{chakraborty2025design}
reports approximately $11.68~\mathrm{mW}$
of DC heater power per ring to tune a
GF45CLO resonator over half of its free
spectral range. This value characterizes
that tuning range and is not the measured
holding overhead of the compared pLatch.
The proposed pVOM avoids cavity-resonance
alignment by using a nonresonant storage core.

The programmable latch
in~\cite{ashtiani2025programmable}
also provides complementary outputs,
but uses microring-based universal optical
logic gates. Its projected response time
of $30~\mathrm{ps}$ is distinct from a
demonstrated write-pulse duration.
The distinction of pVOM is its nonresonant
storage core and optical readout through
the existing holding path.

The phase-change memory
in~\cite{rios2015integrated} provides
nonvolatile, multilevel storage, with a
minimum reported optical switching energy
of $13.4~\mathrm{pJ}$ at $0.7\%$
readout contrast.
In comparison, pVOM provides binary
volatile storage. Its incremental write
energy must therefore be considered
together with holding power and retention
duration when comparing these memory classes.

\begin{table}[htbp]
    \centering
    \caption{
    Comparison of photonic memory implementations.
    }
    \label{tab:memory_comparison}
    \small
    \setlength{\tabcolsep}{5pt}
    \renewcommand{\arraystretch}{1.2}

    \begin{tabular}{lcccc}
        \toprule
        Memory
        & Write pulse
        & Switching
        & Area
        & Resonance \\
        & width
        & energy (pJ)
        & ($\mathrm{mm}^{2}$)
        & alignment \\
        \midrule

        pLatch~\cite{abdullah2025demonstration}
        & $50~\mathrm{ps}$
        & 0.5
        & 0.1
        & Yes \\

        Ashtiani~\cite{ashtiani2025programmable}
        & --
        & 0.4$^{a}$
        & --
        & Yes \\

        Rios et al.~\cite{rios2015integrated}
        & $10~\mathrm{ns}$
        & 13.4$^{b}$
        & --
        & No \\

        This work
        & $10~\mathrm{ps}$
        & 0.562$^{c}$
        & 0.0325
        & No \\

        \bottomrule
    \end{tabular}

    \vspace{5pt}

    \begin{minipage}{0.95\linewidth}
        \footnotesize
        $^{a}$Projected energy; a response time of
        $30~\mathrm{ps}$ is estimated, but a
        write-pulse duration is not reported.\\
        $^{b}$Minimum optical switching energy
        at $0.7\%$ readout contrast.\\
        $^{c}$Incremental electrical and
        source-referred laser energy at 20\%
        wall-plug efficiency; holding energy
        excluded.\\
        Pulse widths do not specify complete
        write-cycle times. Energy values follow
        the accounting used in each work.
    \end{minipage}
\end{table}

\subsection{Array Operation}

\begin{figure}[htbp]
    \centering
    \includegraphics[width=0.65\linewidth]
    {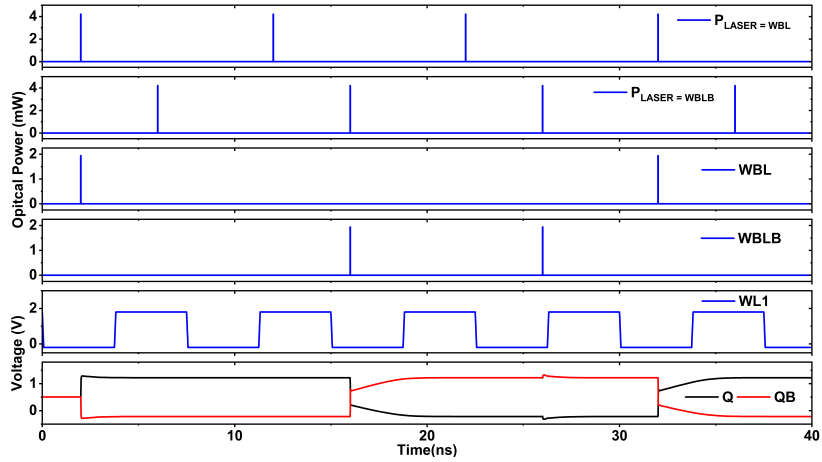}
    \caption{
    Simulated write access for one bitcell
    in a $4\times4$ pVOM array.
    Optical write pulses pass through the
    access VOAs when WL is low and are
    suppressed when WL is high, preserving
    the stored state.
    }
    \label{fig:array_write}
\end{figure}

A $4\times4$ array is simulated using the
architecture in Fig.~\ref{fig:pvom_cell}(c).
Fig.~\ref{fig:array_write} shows the source
pulses, transmitted write signals,
wordline voltage, and storage-node response
of one bitcell.

With WL low, the WBL pulse at approximately
$2~\mathrm{ns}$ sets $Q$ high, and the
WBLB pulse at approximately
$16~\mathrm{ns}$ reverses the stored state.
Source pulses arriving while WL is high
are suppressed by the access VOAs and
do not change the stored logic state.
The WBLB pulse near $26~\mathrm{ns}$
reinforces the existing state, producing
only a transient, whereas the WBL pulse
near $32~\mathrm{ns}$ switches the cell again.

These results demonstrate selective writing
and retention during unselected intervals
for the observed cell.
The nonresonant core simplifies wavelength
management in larger arrays, although optical
distribution losses, access leakage, and
device variations remain scaling considerations.
Independent row readout also requires
isolation of unselected cells from the
shared optical read outputs.

\FloatBarrier

\section{Conclusion}

We proposed a VOA-based photonic memory using
the GlobalFoundries 45SPCLO platform.
The nonresonant core maintains complementary
states through electro-optical feedback
without cavity-resonance alignment.
The cell retains data as long as optical
and electrical power are supplied.
Circuit simulations demonstrate writing
with $10~\mathrm{ps}$ optical pulses at
$2~\mathrm{mW}$ peak power and verify
retention for $5~\mathrm{s}$ per logic state. Wordline-controlled access supports selective
writing, with operation evaluated in a
$4\times4$ array. Optical readout uses the
existing holding input and output splitters
without additional read VOAs or a separate
laser source.
The evaluated transition requires an
incremental write energy of
$0.562~\mathrm{pJ}$, and the bitcell layout
occupies $0.0325~\mathrm{mm}^{2}$,
providing a foundry-compatible implementation
of volatile photonic memory with optical
write and read access.


\section*{Funding}

This work was supported by the Defense Advanced
Research Projects Agency (DARPA) under Grant
No.~N660012424003, and in part by the DoD ASPEN
program and the NSF under Award
No.~CCF-2319617.

\section*{Disclosures}

The authors declare no conflicts of interest.

\section*{Data Availability}

Data will be made available on request.

\bibliographystyle{unsrt}
\bibliography{sample}

\end{document}